\documentstyle[psfig]{l-aa}

\def\degr{\hbox{$^\circ$}}
\def\mdot {\hbox {$\dot M$}}
\def\vz  {V$_{\rm z}$}
\def\az  {A$_{\rm z}$}
\def\sz  {S$_{\rm z}$}
\def\dz  {D$_{\rm z}$}
\def\szhid  {S$_{\rm z}^{\rm hid}$}
\def\dzhid  {D$_{\rm z}^{\rm hid}$}
\def\sigvz  {$\sigma_{{\rm V}_{\rm z}}$}
\def\sigaz  {$\sigma_{{\rm A}_{\rm z}}$}
\def\cygx {\hbox{Cygnus X-2}}

\def\la{\mathrel{\mathchoice {\vcenter{\offinterlineskip\halign{\hfil
$\displaystyle##$\hfil\cr<\cr\noalign{\vskip1.5pt}\sim\cr}}}
{\vcenter{\offinterlineskip\halign{\hfil$\textstyle##$\hfil\cr<\cr
\noalign{\vskip1.0pt}\sim\cr}}}
{\vcenter{\offinterlineskip\halign{\hfil$\scriptstyle##$\hfil\cr<\cr
\noalign{\vskip0.5pt}\sim\cr}}}
{\vcenter{\offinterlineskip\halign{\hfil$\scriptscriptstyle##$\hfil
\cr<\cr\noalign{\vskip0.5pt}\sim\cr}}}}}
\def\ga{\mathrel{\mathchoice {\vcenter{\offinterlineskip\halign{\hfil
$\displaystyle##$\hfil\cr>\cr\noalign{\vskip1.5pt}\sim\cr}}}
{\vcenter{\offinterlineskip\halign{\hfil$\textstyle##$\hfil\cr>\cr
\noalign{\vskip1.0pt}\sim\cr}}}
{\vcenter{\offinterlineskip\halign{\hfil$\scriptstyle##$\hfil\cr>\cr
\noalign{\vskip0.5pt}\sim\cr}}}
{\vcenter{\offinterlineskip\halign{\hfil$\scriptscriptstyle##$\hfil
\cr>\cr\noalign{\vskip0.5pt}\sim\cr}}}}}

\begin{document}

\thesaurus{06 (02.01.2;  08.09.2;
               08.14.1; 13.25.5)}

\title{Ginga observations of Cygnus X-2 }

\author {R.A.D.~Wijnands\inst{1},\,
         M.~van der Klis\inst{1}\,
         E.~Kuulkers\inst{2}\,
	 K.~Asai\inst{3}\,
 	 G.~Hasinger\inst{4}}

\offprints {R.A.D. Wijnands, e-mail: rudy@astro.uva.nl}

\institute{Astronomical Institute ``Anton Pannekoek'',
           University of Amsterdam
           and Center for High Energy Astrophysica (CHEAF),
	   Kruislaan 403,
           1098 SJ Amsterdam,
           The Netherlands
\and       ESA/ESTEC, Astrophysics Division (SA), P.O. Box 229,  2200 AG Noordwijk, The Netherlands
\and       Institute of Space and Astronautical Science, 
           1-1 Yoshinodai 3-chome, Sagamihara, Kanagawa 229, Japan
\and       Astrophysikalisches Institut Potsdam, An der Sternwarte 16,
           14482 Potsdam, Germany}

\date{Received ; accepted}

\maketitle
\markboth{R.A.D.~Wijnands et al.: Ginga observations of Cygnus~X-2}{}

\begin{abstract}
We have analysed all available X-ray data on the low-mass X-ray binary
Cygnus~X-2 obtained with the Ginga satellite. A detailed analysis of
the spectral and fast timing behaviour of these 4 years of data
provides new insights in the behaviour of this Z source.  We confirm
the previously observed recurrent patterns of behaviour in the X-ray
colour-colour and hardness-intensity diagrams consisting of shifts and
shape changes in the Z track. However, we find a continuous range of
patterns rather than a discrete set. The source behaviour in the
diagrams is correlated with overall intensity, which varied by a
factor of 1.34 in the Ginga data. We find that when the overall
intensity increases, the mean velocity and acceleration of the motion
along the normal branch of the Z track increase, as well as the width
of the normal branch in the hardness-intensity diagram.  Contrary to
previous results we find that, during different observations, when the
source is at the same position in the normal branch of the Z track the
rapid X-ray variability differs significantly.  During the Kuulkers et
al. (1996a) ``medium'' level, a normal branch quasi-periodic
oscillation is detected, which is not seen during the ``high'' overall
intensity level.  Also, during the high overall intensity level
episodes the very-low frequency noise on the lower normal branch is
very strong and steep, whereas during the medium overall intensity
level episodes this noise component at the same position in the Z
track is weak and less steep.  The explanation of the different
overall intensity levels with a precessing accretion disk is difficult
to reconcile with our data.  Furthermore, we found that the frequency of the
horizontal branch quasi-periodic oscillation decreases when Cygnus~X-2
enters the upper normal branch, giving a model dependent upper limit
on the magnetic field strength at the magnetic equator of
$\sim8.5\times10^9$ G.  We also report five bursts, with durations
between two and eight seconds, whose occurrence seems to be
uncorrelated with location in the Z track, overall intensity level or
orbital phase.  The burst properties indicate that they are not
regular type I bursts.

\end{abstract}

\section{Introduction \label{introduction}}

Cygnus~X-2 (discovered by Bowyer et al. 1965) is a very well studied
bright low-mass X-ray binary (LMXB). Hasinger \& van der Klis (1989)
classified the brightest LMXBs into two sub-classes, i.e.  the ``Z''
sources and the ``atoll'' sources, on the basis of their correlated
X-ray spectral and fast timing behaviour. Cygnus~X-2 was classified as
a Z source. In the X-ray colour-colour diagram (CD) Z sources trace
out a Z shaped track and move smoothly, without jumps, through the
branches.  In all the six known Z sources, except GX 349+2, three
branches have been identified, the horizontal branch (HB), the normal
branch (NB) and the flaring branch (FB). The transition between the HB
and the NB is called the hard vertex, and between the NB and the
FB the soft vertex. The variation of one single parameter, i.e. the
mass transfer rate to the compact object (\mdot), is thought to produce the
tracks (e.g. Hasinger \& van der Klis 1989, Lamb 1991). \

Recent studies of EXOSAT data of the Z sources (see Kuulkers et
al. 1994a, 1996a,b; Kuulkers \& van der Klis 1996) indicate that the Z
sources can be divided into two groups, the Cyg-like sources (Cyg~X-2,
GX~5-1 and GX~340+0) and the Sco-like sources (Sco~X-1, GX~349+2 and
GX~17+2). The Cyg-like sources show motion of the Z pattern in the
colour-colour diagram and the hardness-intensity diagram (HID), their
HBs are horizontal, and when the sources are on their short FBs the
X-ray intensity generally decreases. The Sco-like sources, instead, do
not display significant motion of their Z pattern in the CD, their HBs
are almost vertical, and the X-ray intensity increases when they move
onto their extended FBs. It was proposed (see Kuulkers \& van der Klis
1995b, and references therein) that the Sco-like sources are viewed
face-on (low orbital inclination) and the Cyg-like sources more
edge-on (higher orbital inclination), and that the motion of the
Cyg-like sources could be caused by accreting material, such as a
precessing accretion disk getting into the line of sight.

Theoretical studies (Psaltis et al. 1995) indicate that a
difference of magnetic field strength can explain some of the
differences between the Cyg- and the Sco-like sources. In their model,
the Cyg-like sources have a somewhat higher magnetic field strength (B
$\sim 5\times 10^9$ Gauss) than the Sco-like sources (B $\sim 10^9$
Gauss). Their so-called unified model can explain the Z tracks in the
CDs and HIDs and the difference in the Z shapes between the Cyg-like and
Sco-like sources. However, no explanation is given for the fact that
the Cyg-like sources display motion of the tracks, whereas the Sco-like
sources do not.

Cygnus~X-2 shows the most pronounced motion of the Z pattern in the CD
and the HID of all Z sources.  One of the first observations of Z
pattern motion in the HID was found by Vrtilek et al. (1986) using
Einstein MPC data. They observed a factor of two increase in intensity
between different epochs but also an increase of the intensity on time
scales less than a day. At that time the Z track behaviour of
Cygnus~X-2 was not yet known, so they interpreted the variations as
due to orbital effects.  Nowadays we know that most of the intensity
variations of Cygnus~X-2 are due to the non-periodic motion of the
source through the Z track, and most likely associated with changes in
\mdot. However, not all intensity variations that were found by
Vrtilek et al. (1986) can be explained in this manner. In their
Fig.~2, three branches are present which now can probably be
identified as NBs.  These branches are shifted with respect to each
other, making this in retrospect the first indication that Cygnus~X-2
displays motion of the Z-track in the HID.

Clear evidence for Z track motion in the CD and the HIDs was found
when different EXOSAT observations of \cygx~ were compared with
each other (Hasinger et al. 1985b, Hasinger 1987, 1988).  Hasinger
(1987) discussed six EXOSAT observations of Cygnus~X-2. He found that
the Z tracks of different observations were displaced in the HIDs with
respect to each other. Hasinger et al. (1985b) and Hasinger (1988)
reported a state of Cygnus~X-2 with very low count rates. The count
rates, especially in the high energy bands, flared and the spectrum
was hard. In the CD this observation looked like a large FB, and it is
displaced to harder colours with respect to other Cygnus~X-2 EXOSAT
observations (see also Kuulkers et al. 

\begin{table*}[t]
\caption[]{Log of the observations \label{obslog}}
\begin{flushleft}
\begin{tabular}{llllll}
\hline
Nr & start (UT) & end (UT) & total time (h) & mode & binary phase$^a$\\
\hline
1 & 1987-06-07  15:29 & 1987-06-10 21:43 &  6.74  & MPC1, MPC3, PC &
0.70-0.03 \\
2 & 1988-06-10  05:33 & 1988-06-14 04:05 & 28.80  & MPC2, MPC3, PC &
0.14-0.54 \\
3 & 1988-10-06  00:50 & 1988-10-07 23:40 & 18.24  & MPC1, MPC2, MPC3,
PC & 0.12-0.30 \\
4 & 1989-10-22  01:41 & 1989-10-24 01:06 &  8.07  & MPC1, PC &
0.81-0.01 \\
5 & 1990-11-30  16:13 & 1990-12-02 13:31 &  2.32  & MPC1, MPC3  &
0.91-0.11 \\
6 & 1991-05-15  01:16 & 1991-05-16 10:00 &  7.83  & MPC1, MPC3  &
0.71-0.85 \\
7 & 1991-06-09  18:26 & 1991-06-11 06:41 & 11.32  & MPC1, MPC2, MPC3 &
0.33-0.48 \\
\hline
\multicolumn{6}{l}{$^a$ \small orbital phase using the ephemeris given
by Crampton \& Cowley (1980), see Kuulkers et al. (1996a) note 4.}\\
\end{tabular}
\end{flushleft}
\end{table*}

\normalsize

Vritlek et al. (1988) investigated the long-term temporal variability
of Cygnus~X-2 using six different instruments on three satellites (OSO
8, HEAO 1, and Einstein). Three count rate states between 2 and 10 keV
were found. During the lowest state the spectrum was harder than
during the other two states.\

The motion of the Z trough the CD and HID was again reported in early
Ginga observations of Cygnus~X-2 reported by Hasinger et
al. (1990). They found that the Z track in the HID during the October
1988 observation was shifted to higher intensities (by a factor two)
with respect to the July 1988 observation. In the CD the Z track of
the October 1988 observation was shifted downward parallel to the NB
with respect to the July 1988 observation. The shape of the Z track in
both the CD and the HID was different between these two observations.
During the July 1988 observation the Z track performed a loop in the
HID at the NB-FB transition, such that along the FB the source first
rose in intensity and later decreased. During the October 1988
observation no such loop was found. Instead, along the FB the source
immediately started to decrease in intensity.  In the CD of the July
1988 observations the HB is diagonal and the FB is large. In the
October 1988 CD the HB is horizontal and the FB is very small.\

Kuulkers et al. (1996a) reanalysed all EXOSAT observations of
Cygnus~X-2. They also reported motion of the Z pattern and different
shapes of the Z track. They compared their results with all previous
reports on Cygnus~X-2, and described the phenomenolgy in terms of
three different intensity levels (possibly part of a continuous
sequence). In the rare low level Cygnus~X-2 displays a hard spectrum
at the lowest intensities. These episodes seemed to occur at binary
phases 0.8-0.2 (with phase 0.0 defined as the X-ray source superior
conjunction), however not in each orbital cycle.  Vritlek et
al. (1988) and Kuulkers et al. (1996a) discussed various possible
explanations, e.g. that at these times the secondary hides its heated
face and/or part of our view of the inner disk region, so that most of
the observed radiation then comes from a scattering hot corona
surrounding the inner accretion disk, or that a tilted disk or a hot
spot at the outer disk hide much of the inner disk region from our
view.  The other two intensity levels were called the medium and the
high level. A periodicity in the occurence of these intensity levels
could not be found. The shape of the Z track in the HIDs was dependent
on these intensity levels. As the source moves onto the FB during high
level episodes, the intensity immediately starts to decrease and the
FB in the HID is almost horizontal. They called this a
'colour-independent' dip. When the source moves up the FB during
medium level episodes the intensity first increases and then
decreases. The FB in the HID is also pointed upwards and therefore
referred to as a 'colour-dependent' dip. No Z track was observed
during the low level. \

The fast timing behaviour of Z sources is closely related to the
position of the sources in the CD and HID (Hasinger \& van der Klis
1989). On all branches a very-low frequency noise (VLFN) and a high
frequency noise (HFN) component in the power spectra can be
identified. On the horizontal and normal branches also low frequency
noise (LFN) can be identified which decreases in amplitude from the HB
to the upper normal branch and disappears on the lower normal branch.
Quasi-periodic oscillations (QPOs) are found on the horizontal branch
(called HBOs) with a frequency from $\sim$12 Hz at the left end of the
HB up to $\sim$55~Hz near the hard vertex, and then remains constant
upto halfway the normal branch. However, Wijnands et al. (1996a) found
a decrease of the frequency of what is probably the HBO down the NB in
GX~17+2. On the NB QPOs (NBOs) are observed with a frequency between
5--7 Hz.  Sometimes the HBO is seen simultaneously with the NBO,
indicating different mechanisms are responsible for these QPOs.  On
the flaring branch sometimes a QPO (FBO) is seen with a frequency near
7 Hz at the beginning of the FB, but rapidly increasing in frequency
up to $\sim 20$ Hz further up the FB.  The 7-20 Hz flaring branch QPO
has not been found on \cygx.  However, Kuulkers \& van der Klis
(1995a) detected a QPO on the FB of Cygnus~X-2 with a frequency
of 26 Hz. This QPO was found in an intensity dip on the FB and is
believed to be a different phenomenon from the 7-20 Hz FBO.\

Two Z sources are known to show bursts. GX~17+2 (Tawara et al. 1984;
Kahn \& Grindlay 1984; Sztajno et al. 1986; Kuulkers et al. 1994b,
1996b) shows bona fide type I burst (thermonuclear flashes on the
neutron star surface, Hoffman et al. 1978). In Cygnus~X-2 burst-like
events have been reported (Kahn \& Grindlay 1984; Kuulkers et
al. 1995).  Kuulkers et al. (1995) detected 9 burst-like events in the
archival EXOSAT data of Cygnus~X-2. These burst-like events had a
short duration (t$_{\rm burst}\sim $3 s), did not show evidence for
cooling, and did not occur in specific regions of the Z track. To date
it is unclear wether these burst-like events are genuine type I
bursts.\

In this paper we present an uniform analysis of all available data
obtained by the Large Area Counter (LAC) instrument on board the Ginga
satellite.

\section{Observations and Analysis \label{obssz}}

\subsection{The satellite and the observations \label{observations}}

Cygnus~X-2 was observed with the Ginga satellite (Makino and the
ASTRO-C team 1987) during 7 observations. For a log of the
observations we refer to Table \ref{obslog}.  Due to Earth
occultations and the high background in the South Atlantic Anomaly
(SAA) the observations were broken up into many pieces with a length
of several tens of minutes. 
 
The LAC instrument (Turner et al. 1989) on board Ginga was used in
several different observation modes : the MPC1, MPC2, MPC3 and the PC
mode.  In the MPC1 and the MPC2 (hereafter MPC1/2) modes the source
was observed in 48 photon energy channels and the time resolution
ranged from 62.5 ms to 16 s.  In the MPC3 mode the source was observed
in 12 channels with a best high time resolution of 7.8 ms.  In the PC
mode only four overlapping channels were used and the time resolution
was 1 or 2 ms.  By manipulating the coarse gain (CG) amplifier, which
had two levels, the energy range could be set to either 0.5--37 keV
(low CG level) or to 0.3--18.7 keV (high CG level). By changing the
high voltage (HV) levels the energy range could be extended to even
higher energies (e.g. up to 60 keV). Due to these different CG
amplifier settings, HV level settings and lower discriminator settings
(see Turner et al. 1989) different sets of energy channels were used
at different times.

\begin{table*}
\begin{flushleft}
\caption[b]{The energy bands used to create colour-colour and
hardness-intensity diagrams. We used the following definitions of the
colours~: Soft Colour~ = $^{10}$log(Band 2 / Band 1), Hard Colour =
$^{10}$log(Band 3 / Band 2). The intensity is the
count rate in the energy range given in colum 6  \label{bands}}
\begin {tabular}{llllll}
\hline
Observation period  & Mode            & Band 1 & Band 2 
& Band 3  & Energy range used for the total intensity       \\
    & & keV & keV & keV & keV\\
\hline
1987 and  1988      & MPC1/2 and MPC3 & 2.3--4.7  & 4.7--7.0  &
7.0--18.6 & 2.3--18.6  \\
1991                & MPC1/2          & 2.3--4.7  & 4.7--7.0  &
7.0--18.6 & 2.3--18.6  \\
1991                & MPC3            & 2.4--4.7  & 4.7--7.1  &
7.1--18.8 & 2.4--18.8  \\
\hline
1989 and 1990       & MPC1/2          & 2.9--4.8  & 4.8--6.7  &
6.7--19.0 & 2.9--19.0  \\
1989 and 1990$^a$   & MPC1/2 and MPC3 & 1.2--3.8 & 3.8--7.6  &
7.6--19.0 & 1.2--19.0  \\
\hline
\hline
1987                & PC              & 0.6--2.9  & 2.9--9.1  &
7.9--12.3~ & 0.6--9.1 + 0.8--12.3 \\
1988                & PC              & 0.8--2.9  & 2.9--9.1  &
7.9--12.3~ & 0.8--9.1 + 0.8--12.3 \\
\hline
1989                & PC              & 1.2--5.7  & 1.2--15.7  &
5.7--17.9~ & 1.2--17.9 + 1.2--15.7 \\
\hline
\multicolumn{6}{l}{$^a$ \small These bands were used in order to determine in which part of the Z Cyg X-2 was during the 1990 MPC3 observation (see text).\normalsize}\\
\end{tabular}
\end{flushleft}
\end{table*}

Generally the satellite attitude control system kept the viewing
direction within 30 arcmin from the position of \cygx, which
corresponds to a collimator transmission of $\ga$ 70\%. The systematic
error in the observed count rates due to uncertainties in the
collimator transmission is of the order of a few precent (Hertz et
al. 1992). Due to, e.g., a long period without attitude control
measurements the offset angle sometimes increased causing the
collimator transmission to decrease to about 40\%.  The LAC collimator
shows energy-dependent transmission effects due to reflection of soft
X-rays (below $\sim$6~keV) from the collimator walls at large offset
angles (Turner et al. 1989). This low energy reflection was not taken
into account when the intensities were corrected for collimator
transmission (Sect. \ref{makingCDHID}), which gives rise to an
overestimation of the count rates for the lowest energy channels. When
we encountered data with a collimator transmission below $\sim$60 \%
we used Figs. 13 and 14 of Turner et al. (1989) in order to correct
the count rates below 6 keV. Usually this means that the count rates
were corrected for an overestimation of 3--4\%. Sometimes the offset
angle was even higher and the collimator transmission came below
$\sim$20\%. At such high offset angle the collimator response is
independent of incident photon energy (Turner et al. 1989). However,
at these times the fractional uncertainty in the collimator
transmission, and thus in the corrected count rates, due to aspect
jitter is $\ga$ 15\% (Hertz et al. 1992). For this reason, we did not
use any data with a collimator transmission $\la 40$\% in our
analysis. \

\begin{figure}[t]
\psfig{figure=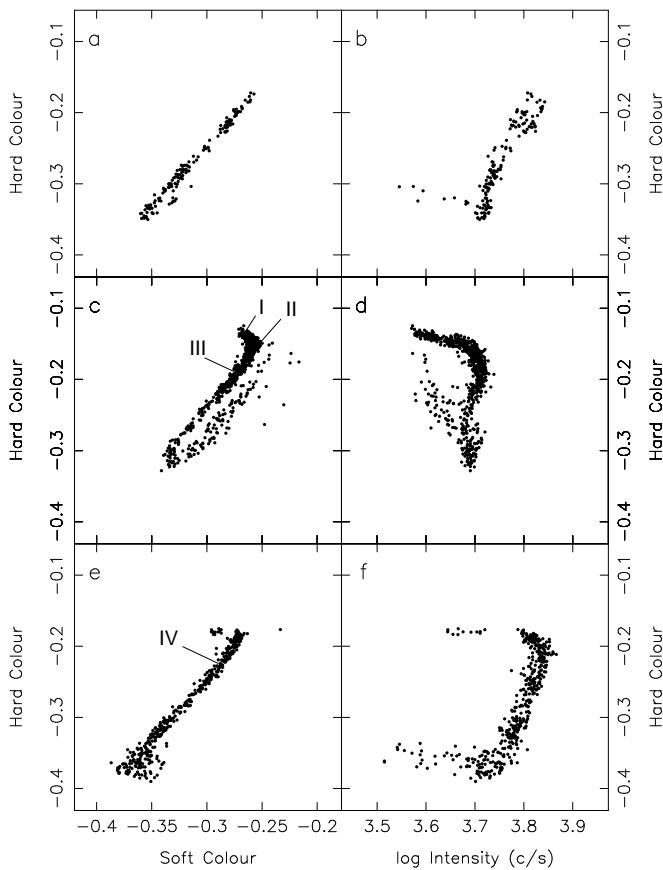,width=88mm}
\caption[]{\label{fig1}{\em The June 1987, June 1988, and October 1988
observations}. MPC modes: {\bf a}, {\bf c}, and {\bf e} are the
colour-colour diagrams and {\bf b}, {\bf d}, and {\bf f} are the
hardness-intensity diagrams for the June 1987, June 1988 and October
1988 observation, respectively. For the definitions of the colours and
the intensity see text and Table 2. The position of the I, II and the
II ({\bf c}) indicate the position in the Z track were the three June
1988 burst occurred. The position of the IV ({\bf e}) indicates were
in the Z the October 1988 burst occurred}
\end{figure}

\begin{figure}[t]
\psfig{figure=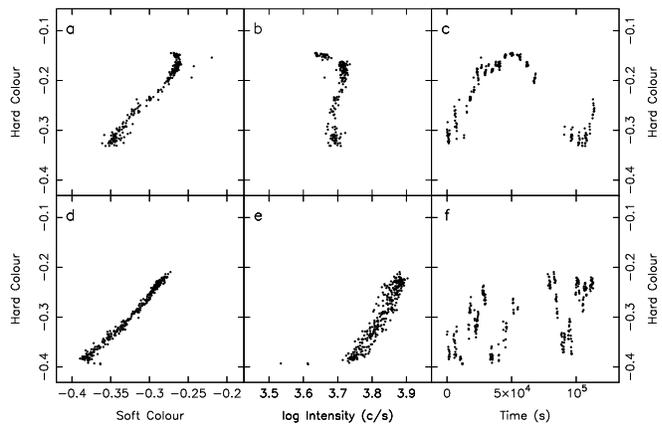,width=88mm}
\caption[]{\label{fig2}{\em The May 1991 and June 1991
observations}. MPC modes: {\bf a} and {\bf d} are the colour-colour
diagrams, {\bf b} and {\bf e} are the hardness-intensity diagrams, and
{\bf c} and {\bf f} are the hard colour curves for the May and June
observations, respectively}
\end{figure}

\subsection{Colour-colour and hardness-intensity diagram analysis \label{makingCDHID}}

For our studies of the CDs and the HIDs we used data rebinned to a
time resolution of 4 seconds, which were background substracted and
corrected for deadtime and aspect using the ISAS analysis system.  The
PC binned data were not corrected for background because no good
background determinations are available for them.  The 16 s resolution
MPC1 data of the November 1990 and June 1991 observations had multiple
register overflows and were left outside our analysis.  The 4 s
resolution MPC1 data of the June 1987, the October 1989, the May/June
1991 observations and the 2 s resolution MPC2 data of the June/October
1988 and the June 1991 observations had single overflows and the true
count rates could be recovered.\

In order to make CDs and HIDs we calculated the soft colour, the hard
colour and the total intensity. The definitions of those quantities
and the energy bands used in order to calculated them, are given in
Table \ref{bands}.  Instead of the colours and intensities, usually
used in the literature, we used the logarithmic value of those
quantities. Any process which influences the intensity in the
specified energy bands by multiplication, i.e. the instrumental
response, produces only a shift in the CD and the HIDs. Using
logarithmic values of the colours and the intensity therefore makes it 
easier to compare data of the same source obtained by different
instruments, by simply shifting the diagrams. Also, the
\sz~parametrization (Sect. \ref{sz}) in this way does not depend on
the actual values of the quantities or the count rates, but only on
their changes. It can now be used not only for the CD but for the HIDs
as well.\

Usually the total intensity was taken as the sum of the count rates in
the same energy bands as were used to calculate the colours. However,
for the PC data of the 1987 and the 1988 observations the total
intensity was defined as the count rate in all four energy
channels. For the 1989 PC data the count rates in the three lowest
energy channels were summed, as the fourth channel only contained
background photons.  The lowest energy boundary of the 1987 PC data
differs slightly from the lowest boundary in the 1988 PC data
(Table~\ref{bands}). The effect on the Z in the CD and HID is
negligible and therefore we discuss the 1987 and 1988 PC data
together. The 1989 PC data is discussed seperately because of the
different energy boundaries (see below).

In order to compare the MPC1/2 data with the MPC3 data we rebinned the
MPC1/2 data to 12 channels using the same energy boundaries as the
MPC3 data. Although in the 1991 observations the MPC3 data had
slightly different energy boundaries compared to the MPC1/2 data
(Table \ref{bands}), the difference was small and had no significant
effect on the place of the Z in the CD and HID. \

Due to the HV and CG levels of 1989 and 1990 observations (energy
range 1-60 keV) the energy boundaries of their spectral channels
differ from those of the other five observations.  Due to the broad
energy channels the energy boundaries of these observations could not
be succesfully interpolated to the boundaries of the others. Therefore,
we discuss the 1989 and 1990 observations seperately from the other
observations.  In order to get an approximately uniform analysis we
took for the MPC1 mode data colours which resembles the colours used
in the other observations. \

\subsection{\sz~ parametrization \label{sz}}

Hasinger et al. (1990) introduced the concept of rank number of the Z
track in the CD. This concept was refined by Hertz et al. (1992) and
by Dieters \& van der Klis (1996). Here we use the method described by
Dieters \& van der Klis (1996) for the measurement of the position
along the Z, with the difference that we perform all operations on the
logarithmic colours and intensity values (see
Sect.~\ref{makingCDHID}). We therefore transform the two logarithmic
colour coordinates (hard colour and soft colour) into the coordinates
\sz~ (the distance along the Z track) and \dz~ (the distance
perpendicular to the Z track). We also calculated the velocity (\vz)
and the acceleration (\az) along the Z track, as defined as
\begin{equation}
{\rm V}_{\rm Z}(i) = [{\rm S}_{\rm Z}(i+1) - {\rm S}_{\rm Z}(i -
1)]/[{\rm T}(i+1) - {\rm T}(i-1)],
\end{equation}
where \sz(i) is the
position in the Z track on time T(i), and
\begin{equation}
{\rm A}_{\rm Z}(i) = [{\rm V}_{\rm Z}(i+1) - {\rm V}_{\rm Z}(i -
1)]/[{\rm T}(i+1) - {\rm T}(i-1)],
\end{equation}
respectively (see also Dieters \& van der Klis (1996)). The length of
the NB is scaled to 1 and positions on the HB and FB are normalised to
the length of the NB.  The hard vertex corresponds to \sz = 1, and the
soft vertex to \sz = 2. We applied the \sz~parametrization also on the
Z tracks in the HID. We obtained a different set of \sz~ and \dz~
values which we call \szhid~ and \dzhid.\

The distribution of \dz~ or \dzhid~ is a measure for the thickness of
the branches in the CD or HID, respectively. To quantify the thickness
of the branches we calculated the sample standard deviation of the
\dz~ and \dzhid~ distributions using the method discribed in the
Appendix. As a measure of the overall velocity and acceleration in the
CD the standard deviations of the \vz~ and \az~ distribution were
used, again using the method outlined in the Appendix.

\subsection{Power-spectral analysis}

For the power-spectral analysis we used the MPC3 and PC data. We made
FFTs of 128s data segments which resulted a frequency range of
0.0078--64 Hz (MPC3 7.8 ms resolution data) or 0.0078--256 Hz (PC 1.95
ms resolution data) in the power spectra. We calculated the FFTs for
the energy interval 1.2--18.6 keV (June 1987, June/October 1988 and
May/June 1991 MPC3 observations) or 1.2--19.0 keV (November 1990 MPC3
observation). For the PC mode we used the sum of the energy intervals
0.8--12.3 keV and 0.6--9.1 keV (June 1987), or 0.8--12.3~keV and 0.8--9.1
keV (June and October 1988), or 1.2--17.9 keV and 1.2--15.7 keV
(October 1989).  \

The average level of the photon counting noise, modified by deadtime
processes (the Poisson level), was estimated and subtracted using a
counter deadtime of 206 $\mu$s (MPC3 data) and 16.5 $\mu$s (PC data)
(see Mitsuda \& Dotani 1989). In the resulting power spectra several
components can be identified :
\begin{itemize}
\item very low frequency noise (VLFN), which we fitted with $A_{\rm
V} \nu^{-\alpha_{\rm V}}$, where $\nu$ is the frequency, $\alpha_{\rm
V}$ is the power-law index and $A_{\rm V}$ the normalization constant.
\item low frequency noise (LFN) and a high frequency noise (HFN), are
fitted with $A_{\rm L,H} \nu^{- \alpha_{\rm L,H}} e^{- \nu / \nu_{\rm
L,H}}$, where $\alpha_{\rm  L,H}$ is the power-law index, $\nu_{\rm
L,H}$ the cut-off frequency, and $A_{\rm L,H}$ the normalization
constant, for the LFN (L) and HFN (H), respectively.
\item HBO, its harmonic, NBO and FBO, are fitted with Lorentzians :
$A_{\rm Q} {1\over (\nu - \nu_{\rm C})^2 + (\Delta\nu / 2)^2}$, where
$\nu_{\rm C}$ the centroid frequency, $\Delta \nu$ the full width at
half maximum (FWHM) of the QPO and $A_{\rm Q}$ the normalization constant.
\end{itemize}
The fractional rms amplitudes of the various noise components were
determined by integrating their contribution over the following
frequency ranges ; VLFN: 0.001--1 Hz, LFN: 0.01--100 Hz, HFN:
0.01--100 Hz. We note that variations in the collimator transmission
could influences the VLFN (see e.g. the EXOSAT data of GX 17+2
[Kuulkers et al. 1996b]). The errors in the parameter values were
calculated using an error scan through $\chi^2$ space using
$\Delta\chi^2=1$.  Due to the poor statistics it was not always
possible to fit a HFN component. When HFN was fitted we used
$\alpha_{\rm H} = 0$ because in Z sources this index was found to be
consistent with zero (Hasinger \& van der Klis 1989, Dieters \& van
der Klis 1996).  The best way to examine the correlations between the
timing behaviour and the position on the Z track is to select a small
part of the Z track, using the
\sz~ parametrization and determine what the corresponding timing
properties are for that segment. However, our data did not allow this
procedure as we did not have enough data at high time resolution for
this procedure to give sufficient statistics. Therefore we first
selected parts of the data that covered a relatively small \sz~ range
and afterwards determined an average \sz ~ value and range for the
each data set.\

\begin{figure}[t]
\psfig{figure=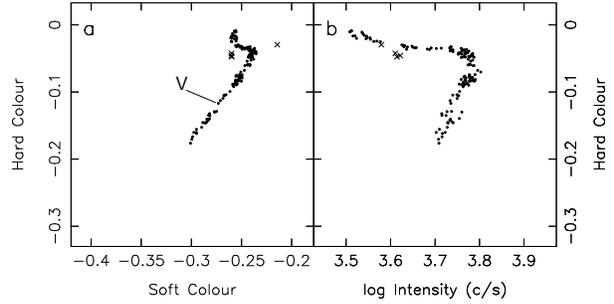,width=80mm}
\caption[]{\label{fig3}{\em The October 1989 and November 1990
observations}. MPC1 modes: {\bf a} is the colour-colour diagram
and {\bf b} is the hardness-intensity diagram.  The dots
({$\bullet$}) are the October 1989 data, the crosses (x) are the
November 1990 data.  The position of the V indicates where in the Z
the October 1989 burst occured}
\end{figure}

\section{Results : the spectral variations}

The CDs and HIDs for all observations are presented in
Figs.~\ref{fig1}--\ref{fig89Oktober_pc}. Each point represents a
time interval of 120 seconds of which $>50$\% contained data. 

\subsection{The 1987, 1988 and 1991 observations}

\subsubsection{June 1987}

In both the CD and HID of the MPC-data (Figs.~\ref{fig1}a and b)
a NB can be seen. No HB is present. The FB is only just seen in the
CD, but it is clearly visible in the HID. The count rate decreases as
soon as the source enters this FB. In the HID, the upper NB is
approximately twice as broad as the middle and lower NB. This
broadening of the upper NB is not seen in the CD.  The PC data
(Figs.~\ref{fig87_88_pc}a and b) only show a NB.  The June 1987 data were
previously discussed by Mitsuda \& Dotani (1989), however, they used
16-second data points. Due to the scatter induced by Poisson
statistics they could not detect the broadening of the upper NB in the
HID.

\subsubsection{June 1988}

This observation was already discussed by Hasinger et al. (1990).
Parts of the data (from 1988 June 12 02:16 to 1988 June 13 08:49 UT
and from 1988 June 14 01:29 to 03:59 UT) were obtained when the offset
angle was about 1\degr, giving a collimator transmission of
40-60\%. We corrected these data for the overestimation of the count
rates in the low photon energy bands (Sect.~\ref{observations}).  In
the CDs of the MPC and PC data (Figs.~\ref{fig1}c and
\ref{fig87_88_pc}c, respectively) clearly a HB, a NB and an extended
FB are visible. The horizontal branch is not horizontal but
diagonal. The HIDs of the MPC and PC data (Figs.~\ref{fig1}d and
\ref{fig87_88_pc}d, respectively) also show all three branches. When
the source moves into the FB, the count rate first slightly increases
and then decreases. \

\subsubsection {October 1988}

The CD and HID of the MPC data (Figs.~\ref{fig1}e and f) were reported
before by Hasinger et al. (1990).  In the CD and HID all three
branches are visible. The HB is really horizontal and when the source
enters the FB the count rate decreases immediately.  In the CD and HID
of the PC data (Figs.~\ref{fig87_88_pc}e and f, not previously reported)
three seperate areas of points are visible.  Comparing these figures
with the MPC data of this observation, the upper and the middle area
(indicated by HV and SV, respectively) are probably near the hard and
soft vertices, respectively. This is confirmed by the fast timing
analysis (see Sect.~\ref{results_timing}).  The status of the third,
lowest area (indicated by E) is unclear.  The fast timing properties
(see Sect.~\ref{results_timing} and Table 5) suggest that
Cygnus~X-2 was near the soft vertex, either on the NB, or on the FB.

\subsubsection {May 1991}

\begin{figure}[t]
\psfig{figure=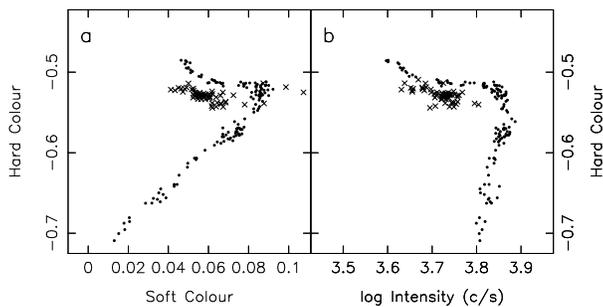,width=80mm}
\caption[]{\label{fig89_90_all_mpc}{\em The October 1989 and November
1990 observation}. MPC1 and MPC3 modes: {\bf a} is the colour-colour
diagram and {\bf b} is the hardness-intensity diagram.  The dots
({$\bullet$}) are the October 1989 data, the crosses (x) are the
November 1990 data. The definitions of colours differ from the ones
used in Fig.~\ref{fig3}, see text and Table 2}
\end{figure}

Part of the data of this observation (from 1991 May 15 01:16 to 04:53
UT) was obtained when the offset angle was large and the collimator
transmission low (40--60\%). We attempted a correction to the low
photon energy channels for the count rate overestimation
(Sect.~\ref{observations}). During the time of low
transmission, Cygnus~X-2 seemed to be on the NB, as judged from the
CD, but in the HID the same data points were displaced from the
NB. Correcting the count rates for the low-energy flux overestimation
was not enough to place the points exactly on the NB in the HID. In
order to show more clearly the soft vertex and the beginning of the FB
(see below) we did not include these points in the HID, and did not
use them in our further analysis.  In the CD (Fig.~\ref{fig2}a)
the HB is short and diagonal. No clear FB can be seen, but a full NB
can.  In the HID (Fig.~\ref{fig2}b) the beginning of a FB can be
seen, which is not made up of the points with the low collimator
transmission. The count rate increases when the source enters the FB.\

\subsubsection {June 1991}

During part of this observation (from 1991 June 10 10:06 to 16:02 UT)
Cygnus~X-2 was observed with a very large offset angle and a
collimator transmission of 10--20\%. We did not include these data in
our analysis because of the large uncertainties in the count rates
(Sect.~\ref{observations}).  The remaining data
(Figs.~\ref{fig2}d and e) show a clear NB, and a hint for a FB, which is
most clearly seen in the HID. When examining the data at higher time
resolution we see a more developed FB. No HB is observed. When the
source moves onto the FB the count rate decreases immediately. The FB
is approximately horizontal. Notice that the very broad NB in the HID
corresponds to a narrow NB in the CD.\

\begin{figure}[t]
\psfig{figure=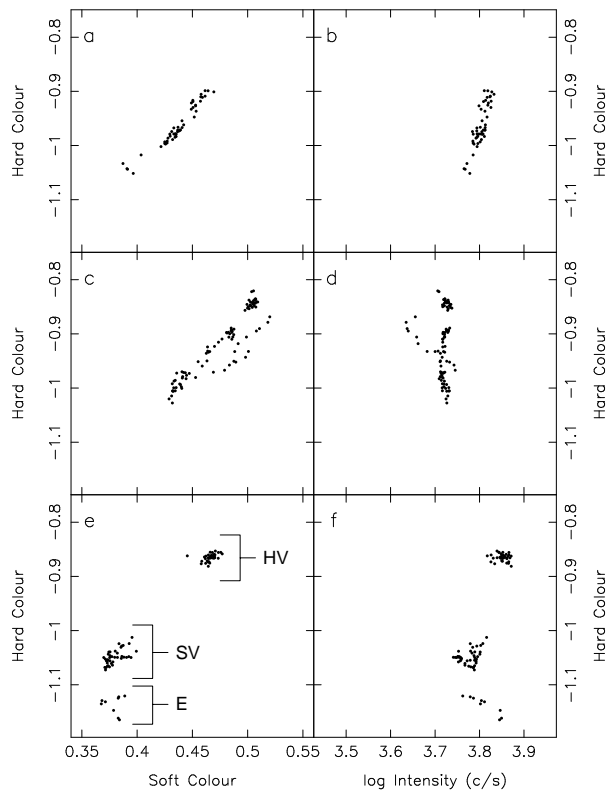,width=80mm}
\caption[]{\label{fig87_88_pc}{\em The June 1987, June 1988 and
October 1988 observations}.  PC mode: {\bf a}, {\bf c}, and {\bf e}
are the colour-colour diagrams and {\bf b}, {\bf d}, and {\bf f} are
the hardness-intensity diagrams for the June 1987, June 1988 and
October 1988 observation, respectively. Most likely the parts of the Z
in the CD of the October 1988 observation ({\bf e}) indicated by HV
and SV belong to the hard vertex and soft vertex, respectively. The
state of the area indicated by E is unclear. See further text }
\end{figure}

\subsubsection {Comparison of the 1987, 1988 and 1991 observations \label{compare}}

Comparing the five observations we find the following. The overall
intensity level, defined as the mean count rate on the NB, changed by
a factor of 1.34 between the observations. The lowest intensities were
observed in June 1988 and May 1991, the highest intensities in October
1988 and June 1991. Intermediate intensities were observed in June
1987. Following the classification of Kuulkers et al. (1996a) our
lowest overall intensities correspond to their ``medium level'' and
our highest overall intensities to their ``high level''. We shall
adopt this terminology in what follows.  As we also observe a level
intermediate between these two, it is obvious from our data that the
medium and high levels are probably part of a continuous range instead
of a discrete set. We did not observe Cygnus~X-2 when it was in the
so-called ``low level'' (Kuulkers et al. 1996a).  We confirm the
previously reported (Hasinger et al. 1990; Kuulkers et al. 1996a and
references therein) correlation between the shape of the Z in the CDs
and HIDs and the overall intensity level. \

\begin{figure}[t]
\psfig{figure=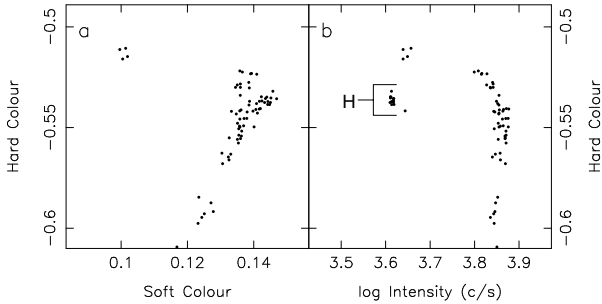,width=80mm}
\caption[]{\label{fig89Oktober_pc}{\em October 1989 observation}. PC
mode: {\bf a} is the colour-colour diagram and {\bf b} is the
hardness-intensity diagram. An harmonic of the HBO was seen in the
part of the Z in the HID ({\bf b}) which is indicated by the H. See also
text}
\end{figure}

In the {\it medium level} the HB is not horizontal but oriented under
an $\sim$45\degr angle in the CD. It is less steep in the HID. In the
CD the HB is quite short compared to the NB. In the HID the HB and NB
are of the same length. The NB is nearly exactly vertical in the HID,
with almost no changes in intensity. The FB is very well developed in
the CD, with large colour changes. In the HID, when the source enters
the FB the count rate first increases, then, further up the FB,
decreases.  In the {\it high level} the HBs in both CD and HID are
nearly horizontal. In the CD the HB is again short compared to the NB,
while in the HID it is approximately the same length as the NB. The NB
in the HID shows a positive correlation between hard colour and
intensity. The FB in the CD is hardly visible, with hardly any colour
changes. In the HID the FB is well developed and horizontal (little
colour changes), and when the source moves onto the FB the count rate
immediately decreases.  The {\it ``intermediate'' level} shows
characteristics of both the medium and the high level. Due to lack of
data nothing can be said about the orientation of the HB during the
intermediate level. The hard colour of the NB in the HID shows a
positive correlation with intensity, similar to the high level,
although not as clear. The FB in the CD is hardly visible, with small
colour changes, also similar to the high level. When the source moves
on the FB in the HID, which is not horizontal, this is similar to the
medium level, the count rate immediately decreases (like in the high
level). In the CD and HIDs, the Z-track shifts downward (to softer
colours), parallel to the NB in the CD, when the overall intensity
level increases.  In the medium level it is at its highest position
(hard colour), in the intermediate level it is lower, and in the high
level it is lowest.\

When we compare the NBs in the HIDs we see that the branch width
changes with overall intensity. In the high level the NB is broader
than in the medium level. The NB in the intermediate level is only
broad in its upper part. Comparing the hard colour curves
(Figs.~\ref{fig2}c and f), we see that the
character of the motion through the NB depends on intensity level. In
the medium level the colours change slower than in the high
level. These two new results are discussed in the next section.

\begin{figure}[t]
\psfig{figure=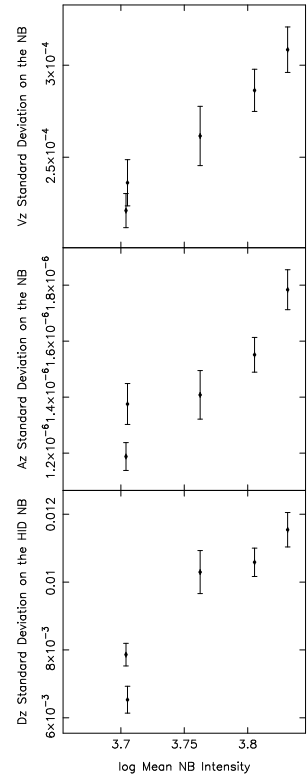,width=65mm}
\caption[]{\label{standevs}{Standard deviation of \vz ~ ({\it upper}),
\az~ ({\it middle}) and \dzhid~ ({\it lower}) on the
NB against the log mean intensity (c/s) on the NB during the different
observations}}
\end{figure}

\subsubsection {The motion through the Z using \sz}

The \sz~ parametrization makes it possible to investigate the
kinematics of the motion along the Z track in more detail (see Dieters
\& van der Klis 1996).  When examining \sz~ as a function of time it
is evident that the source does not jump through the branches, but
moves smoothly along the Z track.  In the high level Cygnus~X-2 seems
to move faster up and down the normal branch than when the overall
intensities are lower (compare Figs.~\ref{fig2}c and f). In order to
investigate this we calculated the velocity (\vz) and acceleration
(\az) distributions (Sect.~\ref{sz}; Dieters \& van der Klis 1996). We
find that the distributions are symmetric about zero : there is no
difference in the motion up and down the Z track. The scatter of \vz~
and \az~ increases from the HB, through the NB, to the FB indicating
that the source moves through the Z most slowly on the HB, faster on
the NB and fastest on the FB.  On the NB the scatter of \vz~ and \az~
increases when the overall intensity increases. In order to quantify
this we determined the standard deviations of \vz~ (\sigvz) and \az~
(\sigaz).

On the NB, \sigvz~ and \sigaz~ show a strong correlation with overall
intensity (see Figs.~\ref{standevs}a and b).  They increase by a factor
$\sim$1.4 and $\sim$1.5, respectively, when the overall intensity
increases by a factor of 1.34.  No correlations between \sigvz~ and
\sigaz~ on the FB and the overall intensity were found. For the HB
the data do not allow a conclusion on this point. 

In order to examine the width of the NB in the HID we applied the \sz~
parametrization on the Z track in the HID (Sect.~\ref{sz}). The
\dzhid~ time series are symmetric around zero.  The lack of points on
the HBs and FBs makes a definitive conclusion about the evolution of
the width of the Z from the HB, via the NB to the FB,
impossible. However, when two or more branches are present the Z width
seems to increase from the HB to the NB, and further on the FB. As
mentioned earlier (Sect.~\ref{compare}) we find that the NB width in
the HID increases when the overall intensity increases. This effect is
visible as a strong increase in the standard deviation of \dzhid on
the NB, with overall intensity (Fig.~\ref{standevs}c).  When the
overall intensity increases by a factor $\sim$1.34, the standard
deviation of \dzhid~ on the NB increases by a factor $\sim$1.75.  On
the other branches the standard deviation of \dzhid~ stays
approximately the same at all overall intensities. When we examine the
NB width in the CDs we do not find any increase in the standard
deviation of \dz ~ with overall intensity : the NB has approximately
the same width at all intensity levels in the CDs.\

\subsection{The 1989 and the 1990 observations}

\subsubsection{The MPC1 mode data}

In order to be able to compare these observations roughly with the
other observations, and because of the very broad photon energy
channels of the MPC3 mode (see Sect.~\ref{makingCDHID}), we made CDs
and HIDs of the MPC 1 data only.  Because during the November 1990
observation the MPC1 mode was used only very briefly, we combined the
October 1989 (dots) and November 1990 (crosses) data in
Fig.~\ref{fig3}.  In both the CD and the HID the HB and
the NB are clearly visible.  No FB is present.  An upward curve can be
seen at the beginning of the HB during the October 1989 observation.
Such an upward curve of the HB was previously reported by Kuulkers et
al. (1996a) in the 1985 EXOSAT observation on day 301/302.  The colour
points of the November 1990 observation are displaced with respect to
the October 1989 data.

\subsubsection{The MPC3 mode data}

In order to include the MPC3 data and to find out on wich part of the
Z track the source was in during the November 1990 observation, we had
to use different colours, with broader energy bands.  For comparison
we did the same for the October 1989 data.  The energy bands used are
given in Table \ref{bands}.  Figure~\ref{fig89_90_all_mpc} shows a clear
HB and NB for the October 1989 data (dots), as before, and the curve
upwards of the HB, although less pronounced.  The
November 1990 data (crosses) show only a HB which is shifted in colour
with respect to the October 1989 data. No upward curve is visible in
the November 1990 data.

\subsubsection{The PC mode data \label{pcmodes}}

The October 1989 PC data CD and the HID (Fig.~\ref{fig89Oktober_pc})
show a NB, the hard vertex and parts of the HB. In the HID, the points
(indicated by the H) on the HB with the lowest intensity and low hard
colour are not on the HB in the CD. They are placed on the hard vertex
in the CD. The fast timing behaviour would suggest that those points
are at the left end of the HB (see Sect.~\ref{HBO})

\subsubsection{Comparison of the 1989 and 1990 observations to the
other observations}

The 1989 and 1990 observations were performed using different energy
bands as compared to the other observations. Therefore, we can not
directly compare their overall intensity levels. However, if we assume
that the difference in the total count rate between 2.3--18.6 keV and
2.9--19.0 keV is minor we see that the overall intensity of the
October 1989 observation is about the same as for the June 1987
observations. There is no NB available for the November 1990
observation so no direct comparison with the other observations of the
mean NB intensity can be made. However, the shift of the HB during
this observation with respect to the October 1989 HB indicates that
the overall intensity of the November 1990 observation was higher than
the overall intensity of the October 1989 observation.  Therefore, we
conclude that the October 1989 observation was taken at an overall
intensity level in between the medium and high level, and the November
1990 observation during a high level episode.

\section{Results : the fast timing behaviour \label{results_timing}}

The results of the power spectra analysis are shown in Tables 4 and 5.

\begin{figure*}[t]
\psfig{figure=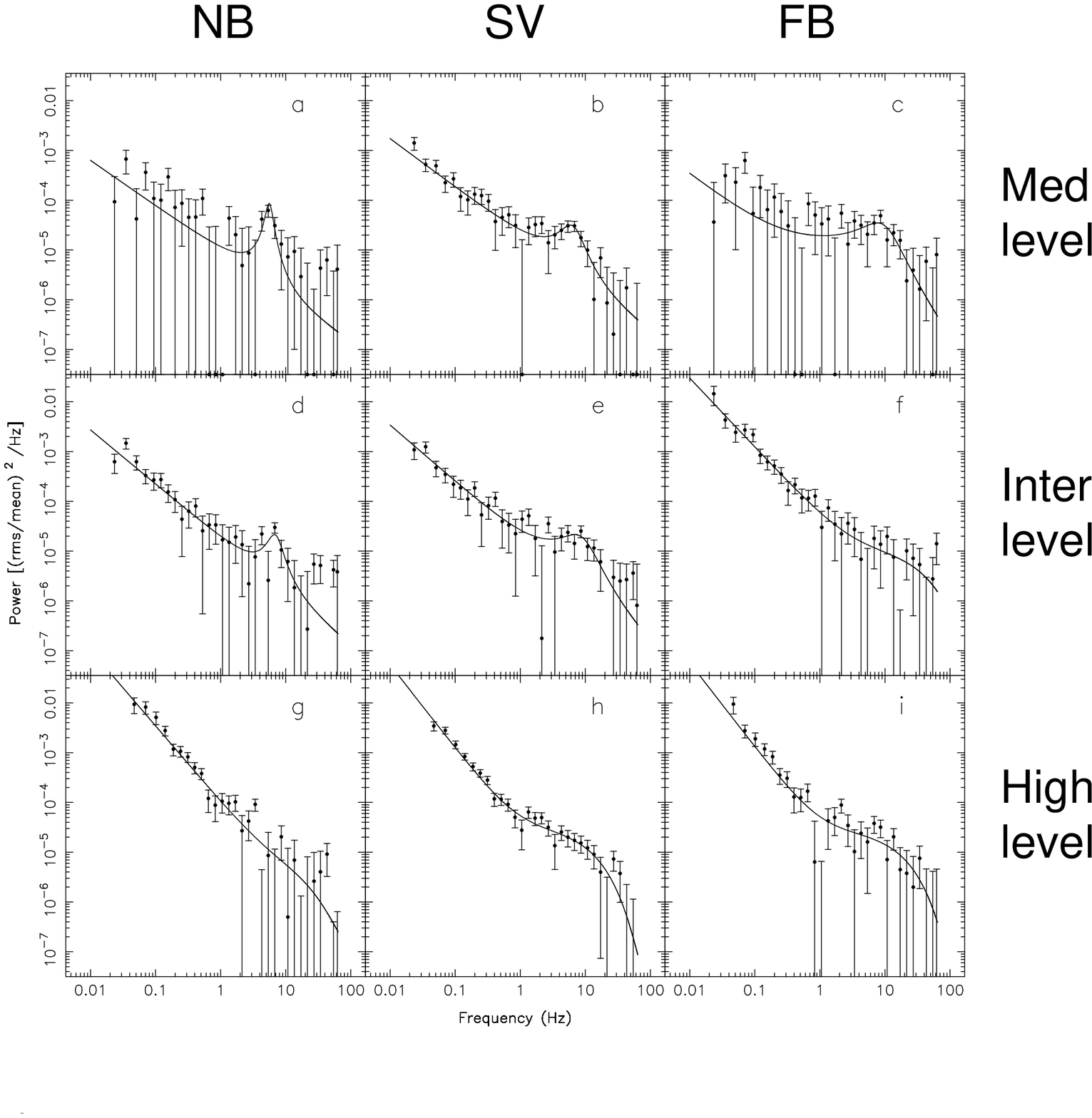,width=150mm}
\caption[]{\label{power_spectra}{Power spectra of different observations
at about the same \sz, the power spectra in the column indicated with
NB, SV and FB, are taken on the lower normal branch, the soft vertex,
and the flaring branch, respectively: {\bf a}, {\bf b} and {\bf c} are
the May 1991 power spectra at \sz =1.79$\pm$0.07, 1.97$\pm$0.10 and
2.05$\pm$0.06, respectively, {\bf d}, {\bf e} and {\bf f} are the June
1987 power spectra at \sz =1.83$\pm$0.02, 1.97$\pm$0.07 and
2.10$\pm$0.03, respectively, and {\bf g}, {\bf h} and {\bf i} are the
October 1988 power spectra at \sz =1.856$\pm$0.008, 2.00$\pm$0.02 and
2.10$\pm$0.01, respectively}}
\end{figure*}

\subsection{Very-low frequency noise (VLFN)}

Since in the HB the VLFN is low (e.g. Hasinger \& van der Klis 1989),
we were in some cases unable to measure the VLFN, at the left end of
the HB. In the June 1987 and June 1988 observations we fixed
the power law index of the VLFN when Cygnus~X-2 is on the HB.

We have high time resolution data of the same part of the Z track
(especially near the soft vertex) but during different overall
intensity levels.  The comparison of these data reveals a very
significant difference of the VLFN properties between the medium and
high overall intensity levels in the lower part of the NB. During the
medium overall intensity observations (June 1988 and May 1991) we
detect a weak (0.5-1\% rms amplitude) and flat (power law index of
$\sim$1) VLFN on the lower NB and near the soft vertex. The rms
amplitude and power law index increase to $\sim$5.3\% and $\sim$1.4,
respectively, on the upper FB in the June 1988 observation. However,
during the high overall intensity observations (October 1988 and June
1991) we detect a very strong (8.5\% and 4.5\% rms, respectively) and
steep (index of 1.4-1.7) VLFN on the lower NB and near the soft
vertex. The VLFN decreases in strength to about 6.7 \% rms
amplitude on the FB in the October 1988 observation. We note that this
strong VLFN near the soft vertex can not be caused by changing
collimator response. During the intermediate overall intensity
observation (June 1987) we see VLFN with a strength of $\sim 1.5$\%
rms and power law index of $\sim$1.1 near the soft vertex, which
increases to $\sim$4.2\% and $\sim$1.4, respectively, on the FB. This
gradual change in the strength and steepness of the VLFN as a function
of overall intensity level is clearly visible in
Figs.~\ref{power_spectra}a-i. This figure shows the power spectra in
the medium level, the intermediate level and the high level at
approximately the same \sz~ values.\

It is remarkable that when Cygnus~X-2 is in the high level the VLFN
near the soft vertex is strong and steep and that when the source is
in the medium level the VLFN at the same \sz~ values is much weaker
and less steep.  During intermediate levels the VLFN is weak and flat
on the NB but strong and steep on the FB.  The correlation of the
amplitude and steepness of the VLFN near the soft vertex with mean
overall intensity level is not strict. At the highest overall
intensities (1991 June) the amplitude is about the half of the one at
the second highest overall intensity (1988 October), and the spectrum
is less steep.\

\subsection{Low-frequency noise (LFN) and high-frequency noise (HFN)}

Due to the scarcity of high time resolution data when \cygx~ was
on the HB or near the hard vertex and the difficulty in determining
\sz, comparisons of the LFN between the observations are
difficult. The June 1987 and 1988 observations show a LFN with an rms
of $\sim$4.7 \% on the HB which decreases to $\sim$3.0 \% on the hard
vertex and upper NB. The November 1990 observation shows that the rms
of the LFN increases from the beginning of the HB to the hard vertex
(from $\sim$5.8 \% to $\sim$6.5 \% rms). The same is seen in the
October 1989 observation (LFN rms from 4.5 \% to 5.5 \%), but in this
observation the rms decreases to 3.8 \% when the source moves into the
upper NB. No clear correlation between the power law index (0.0--0.3)
or the cut-off frequency (4.0-- 13 Hz) with \sz~ is seen. \

When LFN or NBO were fitted it was difficult to fit the HFN
simultaneously, possibly due to interference between them (see also
Kuulkers et al. 1994a).  When it was necessary to fit a HFN component
it had an rms between 1--2\% and cut-off frequencies between 10 and 20
Hz. The October 1989 PC data show a strong HFN ($\sim$ 6\% rms) with a
high cut-off frequency between 30 and 50 Hz.

\subsection{Normal branch QPO (NBO)}

Normal branch QPOs (NBOs) were detected in the medium and intermediate
level observations. To our suprise, none were detected in the high
level observations (see Fig.~\ref{power_spectra}), although we had
high timing data in the lower NB for them. The rms 2$\sigma$ upper
limits ($<0.9$\%) are significantly lower than the rms (1--2.5\%)
of the NBOs in the medium and intermediate levels.  This also shows
that during different overall intensity levels the rapid X-ray
variability at the same position on the Z track differs significantly
from each other.\

When NBOs are seen no clear relation is detected between the amplitude
of the NBO and \sz. However, the FWHM and the centroid frequency seems
to increase from the middle of the NB to the FB. The FWHM increases
from $\sim$2.4 Hz to $\sim$13 Hz and the centroid frequency from
$\sim$5~Hz to $\sim$8 Hz. In the June 1987, the June 1988 and May 1991
observations it was impossible to make a distinction between a broad
NBO or a HFN on the lower FB, although a NBO provided a better fit
(e.g. for the June 1987 observation $\chi^2$/dof = 27.5/29 for the NBO
versus 31.0/30 for the HFN).

\subsection{Horizontal branch QPO (HBO) \label{HBO}}

\begin{figure}[t]
\psfig{figure=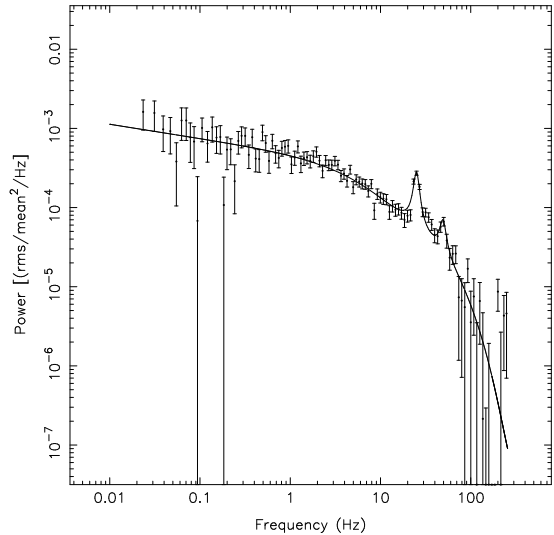,width=80mm}
\caption[]{\label{harmonic}{The power spectrum of the beginning of
the horizontal branch taken in the October 1989 observation. A
harmonic of the HBO is visible at about 50 Hz}}
\end{figure}

Horizontal branch QPOs are seen during all overall intensity levels,
on the HB as well as in the upper NB.  Taking all data into account,
we find that the frequency of the HBO increases from $\sim$31 Hz at
the left end of the HB to $\sim$55 Hz at the hard vertex. The PC data
of the June 1987 and October 1989 observations indicate that when the
source moves down the NB the HBO frequency decreases again to $\sim$
47 Hz. The rms amplitude decreases from the HB to the upper NB, from
$\sim$4\% to $\sim$1\%.

\begin{table*}
\begin{flushleft}
\caption[b]{Bursts \label{bursts}}
\begin{tabular}{llllllllllll}
\hline
No. &Date & Start (UT) & t$_{\rm burst}$ & Mode & Time res. & \sz &
Orbital phase$^{\rm a}$ & $\alpha$ & $\gamma$ & $\tau$ \\
& & & (s) & & (s) & & & & &  (s)\\
\hline
I & 1988-06-12 & 03:56:34 & $\sim$ 8  & MPC2 & 2       & 1.000 & 0.33 & $>$105 & 1.92& 3.66 \\
II & 1988-06-12 & 10:25:52 & $\sim$ 4  & MPC2 & 2       & 0.931 & 0.36 & $>$94 & 3.52 & 2.00 \\
III & 1988-06-13 & 10:47:27 & $\sim$ 8  & MPC2 & 0.0625  & 1.229 & 0.47 & $>$4.1 & 0.63 & 1.51 \\ 
IV & 1988-10-06 & 02:32:25 & $\sim$ 2  & MPC2 & 0.0625  & 1.222 & 0.11& $>$13.3 & 1.74 & 1.09\\
V & 1989-10-22 & 08:04:47 & $\sim$ 4  & MPC1 & 4 &1.4--1.5$^{\rm b}$ & 0.84 &
$>105$& 2.67 & 4.00 \\
\hline
\multicolumn{11}{l}{$^{\rm a}$ \small X-ray phase using the ephemeris given
by Crampton \& Cowley (1980), see Kuulkers et al. (1996a) note 4.}\\
\multicolumn{11}{l}{$^{\rm b}$ \small Only an estimate of \sz, due to the lack of a soft vertex.}\\
\end{tabular}
\end{flushleft}
\end{table*}

As already mentioned in Sect.~\ref{pcmodes}, the PC data of October
1989 show several points at low intensities and high soft
colour. Those points are placed in the CD near the hard vertex but in
the HID near the left end of the HB. The power spectrum corresponding
to these points is shown in Fig.~\ref{harmonic}. A HBO at 25~Hz is
clearly visible (3.8\% rms amplitude) and indicates that Cygnus~X-2
was indeed at the left end of the HB (see Fig.~\ref{fig89Oktober_pc},
the points marked H). A second QPO is visible at $\sim$49~Hz with a
rms amplitude of 2.5\%.  Hasinger et al. (1985a) and Hasinger (1987)
found already evidence for the harmonic of the HBO in the EXOSAT data
of Cygnus~X-2.  However, the frequency ratio was smaller
(1.85$\pm$0.03) than expected for a second harmonic.  Our frequency
ratio (1.96$\pm$0.05) indicates that this feature is indeed the second
harmonic of the HBO.

\section{Results: the bursts \label{results_bursts}}

We found five 2--8 s bursts similar to the burst-like events reported
by Kuulkers et al. (1995). The times and the properties are given in
Table \ref{bursts}.  In Figs. \ref{fig1}c and e and Fig.~\ref{fig3}a
it is indicated where in the Z track the bursts occured. The occurence
of the burst seems to be uncorrelated with the overall intensity
level, the orbital phase, or the location in the Z track (although no
burst were detected on the FB).

For the ratio, $\alpha$, of the average persistent flux to the
time-averaged flux emitted in the bursts (which reflects the ratio
between the gravitational and nuclear burning energy per gram of
accreted matter for thermonuclear bursts), we could only obtain lower
limits (Table \ref{bursts}), since there are many interruptions by SAA
and/or Earth occultations. They were calculated using the average
count rate from the start of the last data gap to just before the
burst. The other typical burst parameters $\gamma$, the ratio of the
mean persistent pre-burst flux and net peak burst flux, and $\tau$,
the ratio of the total integrated net burst flux and the net burst
peak flux (a representation of the burst duration) were also
calculated (see Table~\ref{bursts}). The count rates used in order to
calculate the burst parameters were corrected for deadtime, background
and aspect. Burst IV was observed during time when the satellite was
slewing to the source. Therefore, during that time the collimator
transmission was low ($\la$ 50 \%) and the uncertainty on the count
rate large. The count rate for burst IV was not corrected for the
overestimation of the count rates in the low photon energy bands, due
to the reflection of low energy photons (below 6 keV) agains the
collimator walls (see Sect.~\ref{observations}). 

\begin{figure}[t]
\psfig{figure=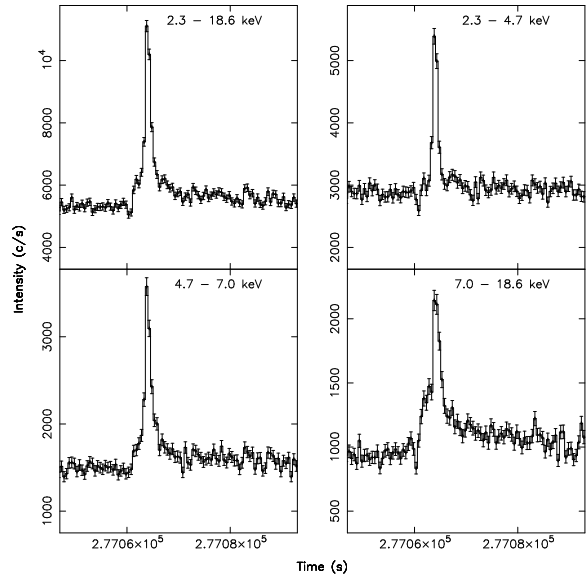,width=80mm}
\caption[]{\label{burst88junimpc2h}{The June 1988 burst (III) in the
MPC2 high bitrate mode, time resolution is 0.5 seconds. Time is given
from the beginning of the observation}}
\end{figure}

\begin{figure}[t]
\psfig{figure=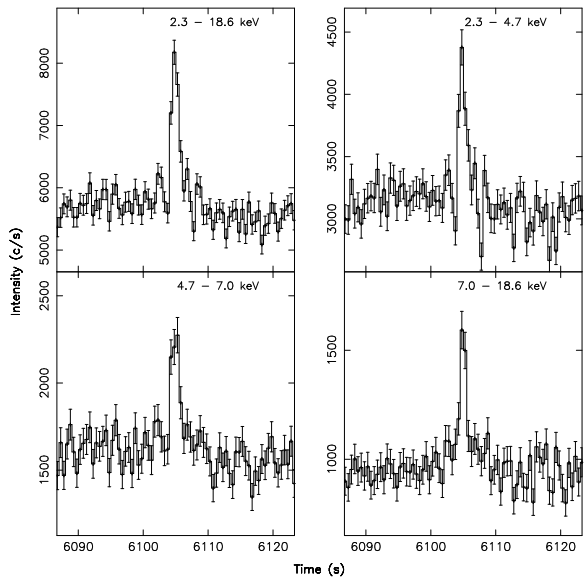,width=80mm}
\caption[]{\label{burst88oktobermpc2h}{The October 1988 burst (IV) in
the MPC2 high bitrate mode, time resolution is 0.5 seconds. Time is
given from the beginning of the observation}}
\end{figure}

Due to the low time resolution of the data obtained during the
occurence of bursts I, II and V, we did not examine the burst profiles
and spectral properties of these events.  The burst profiles, for
different energy bands, of bursts III and IV are shown in
Figs. \ref{burst88junimpc2h} and \ref{burst88oktobermpc2h},
respectively.  These bursts do not show evidence for spectral cooling
as would expected for bona fide type I bursts. Instead, it is clearly
visible that burst III shows evidence for spectral hardening: the
higher the energies, the broader the event. Also, the post-burst count
rate is higher than the pre-burst count rate. This effect is most
prominent at the higher energies. Although these phenomena are not
seen during event IV, no evidence for spectral cooling is seen either.
In the nine bursts in the EXOSAT data of Cygnus~X-2 hints for spectral
hardening were already found (Kuulkers et al. 1995), although the poor
statistics made a definite conclusion impossible. Taking all bursts
into account we conclude that these events are probably not bona fide
type I. We searched for QPOs and periodic oscillations during burst
III and IV and found none. Statistics were insufficient to set meaning
full upper limits.

\section{Discussion \label{discussion}}

We have found several characteristics in the behaviour of Cygnus~X-2
which seem to depend on overall intensity level. In the following we
discuss some of those characteristics in more detail and suggest
possible explanations for the different overall intensity levels and
associated phenomena.

\subsection{The rapid X-ray variability during different overall intensity
levels \label{difference}}

We find conclusive evidence that the rapid X-ray variability at the
same point on the NB differs between different observations.  We find
a different kind of high timing behaviour near the soft vertex when
Cygnus~X-2 is in the medium level and the high level (see
Fig.~\ref{power_spectra}). In the medium level weak, flat VLFN is
detected and a pronounced NBO occurs. In the high level the VLFN
is stronger and steeper, and the NBO is not detectable. In the
intermediate intensity level, the VLFN is strong and steep in the FB,
but weak and flat in the NB, and the NBO is about the same strength as
during the medium level. Our results are the first unambigous
detection of differences in the rapid X-ray variability at a specific
position in the Z track (near the soft vertex) between different
observations (see Dieters \& van der Klis 1996 for a study of Sco X-1
with respect to this issue). So far, we find no evidence for any
difference between the properties of the X-ray variability during
different intensity levels on other parts of the Z track (such as the
HB). However, our data do not allow a very detailed comparison of the
timing behaviour on those parts of the Z track.

It is thought (van der Klis et al. 1985, 1987; Hasinger \& van der
Klis 1989; Lamb 1991) that variations in~\mdot~ are responsible for
the changes in the rapid X-ray variability along the Z and the
formation of the Z track, and that for this reason the rapid X-ray
variability is closely related to the position of the source on the Z
track. However, the shifting and shape changing of the Z track in
Cygnus~X-2 and other Z sources suggest that variations in~\mdot~ can
not explain all aspects of the formation of the Z track. We,
therefore, suggest that both the shape of the Z track and the rapid
X-ray variability are not totally determined by the mass accretion
rate onto the compact object, but also partly by a so far unknown
process, which may also cause the long-term intensity variations (see
Sect.~\ref{longterm}).

Two other Z sources, GX~5-1 and GX~340+0, display motion of their Z
track in the CD (Kuulkers et al. 1994a; Kuulkers \& van der Klis
1996), although not as pronounced as \cygx. Kuulkers et
al. (1994a) and Kuulkers \& van der Klis (1996) did not find any
significant changes in the rapid X-ray varability in GX~5-1 and
GX~340+0, respectively, when the Z track moved through the CD and
HID. If the motion of the Z track in the CD and HID in Cygnus~X-2,
GX~5-1, GX~340+0 are caused by the same phenonemon, we then predict
that the rapid X-ray variability in GX~5-1 and GX~340+0 changes when
the Z tracks of those sources move through the diagram. The reason why
so far no changes have been found is most likely due to the much lower
amplitude of the motion of the Z through the CD and HID of GX~5-1 and
GX~340+0, as compared to \cygx, which suggest that the amplitude
of the difference of the rapid X-ray variability would also be much
smaller.  Moreover, in Cygnus~X-2 sofar only differences in the rapid
X-ray variability were found near the soft vertex and not on e.g. the
HB, while GX~5-1 and GX~340+0 were mainly observed in the HB and upper
NB, and hardly near the soft vertex. Kuulkers et al. (1994a) and
Kuulkers \& van der Klis (1996) could not make a good comparison of
the rapid X-ray variability near the soft vertex when the Z track of
GX~5-1 and GX~340+0, respectively, moved in the CD and HID.

\subsection{Velocity of motion along the Z track}

Van der Klis (1991) suggested that the VLFN could be (partly) due to
motion of the source along the Z track. In our analysis, we found that
both the VLFN fractional amplitude and the velocity of motion along
the Z track increase when the overall intensity level
increases. However, the strongest VLFN was not observed during the
most extreme high level, but during the less extreme, slightly lower
one, whereas the highest velocity was indeed observed during the most
extreme high level. Therefore, we conclude that the motion along the Z
track is at most only partly responsible for the VLFN. As differences
in NB slope in the HID can not explain this, part of the intensity
variations causing the VLFN must take place perpendicular to the NB.

The increase in velocity along the Z track when the overall intensity
increases indicates that the spectrum changes more rapidly during high
level episodes than during medium level episodes.  If the Z track is
traced out by changes in the mass accretion rate onto the neutron
star, then an increase in the velocity of motion along the Z track
indicates that~\mdot~ changes more rapidly when the overall intensity
level increases. However, it then seems unlikely that variations
in~\mdot~ can explain the shifts and changes in shape of the Z track
between intensity levels (see also Sect.~\ref{difference}).  It is
possible that the increase in velocity of motion along the Z track is
due to the same phenomenon causing the Z tracks to shift and the
shapes to change. 

\subsection{The width of the NB in the HID}

The width of the NB in the HID increases when the overall intensity
increases, while the width of the NB in the CD remained approximately
constant. This indicates that the difference in the width of the NB in
the HID is due to intensity variations and not due to spectral
variations. However, at the same time the velocity of motion along the
Z track increases, which indicates an increase in spectral
variations. So, in the HID, when the source moves perpendicular to the
Z track (change in intensity), it also moves along the Z track (change
in spectrum). Perhaps colours are well correlated with \mdot~, but the
intensity varies also due to another (unspecified) process that is
more prominent in the high overall intensity level.

\subsection{The overall intensity variations \label{longterm}}

Several models have been proposed (Priedhorsky \& Holt 1987, and
references therein) in order to explain long-term intensity variations
in X-ray binaries, e.g. long-term variations in \mdot~ and precessing
accretion disks. Precessing neutron stars have also been proposed to
explain the variations (see Priedhorsky \& Holt 1987;
Schwarzenberg-Czerny 1992, and references therein).

\subsubsection{Variations in the mass accretion rate}

Long-term changes in the mass accretion rate have been proposed (see
Priedhorsky \& Holt 1987) to explain the long-term intensity
variations in low-mass X-ray binaries other than Cygnus~X-2 (e.g
4U~1820-30, the Rapid Burster, Aql~X-1, excluding
Her~X-1). However, Kuulkers et al. (1996a) and Wijnands et
al. (1996b) argued that the long-term intensity variations in
Cygnus~X-2 can not be due to variations in the mass accretion
rate. The main argument is that variations in the mass accretion rate
are thought to produce motion of the source along the Z track (see
e.g. Hasinger \& van der Klis 1989), while the Z track is observed
during several different intensity levels.\

\subsubsection{A precessing accretion disk}

In order to explain long term intensity variations, not related to
orbital variations, in high-mass X-ray binaries and Her X-1,
precessing accretion disks have been proposed (see Priedhorsky \&
Holt 1987 and references therein).  The recent detection (Smale et
al. 1996, Wijnands et al. 1996b) of a 78 day period in the RXTE, Vela
5B (see also Smale \& Lochner 1992) and Ariel V all sky monitor data
of Cygnus~X-2 favours a precessing accretion disk in Cygnus~X-2.\

However, explaining the long-term X-ray variations of Cygnus~X-2 with
a precessing accretion disk is not without serious contradictions.  If
we assume that, during the medium overall intensity level, the
emission region is blocked by (part of) the accretion disk and much
radiation is absorbed, scattered and/or reflected, the power spectrum
should be blurred and quasi-periodic oscillations (both the NBO and
the HBO) should be harder to detect than when the emission region is
not blocked by the accretion disk (the high level). However, we see
exactly the opposite: no NBO is detected during high overall
intensity levels. The upper limits are significantly lower than the
actually detected values during the medium levels. If we assume that not
during the medium level but during the high level the emission region
is blocked by the accretion disk, then the difference between the power
spectra are not totally unexpected. However, it is difficult to
explain the increase of the count rate when the accretion disk is
blocking our view of the emission region.\

Another possibility is that the accretion disk blocks, during the
medium overall intensity levels, a localised emission area (e.g. the
neutron star surface), which is not blocked during the high levels. In
this region no NBO occurs and therefore the fractional amplitude of
the NBO during the high level is diluted by the additional
flux. However, not only the strength of the NBO should be affected,
but also the strength of the HBO, except when the HBO would originate
from this localised area, in which case we would expect an inverse
effect. This, however, is unlikely, because during all intensity
levels HBOs are observed at approximately the same fractional
amplitude.\

Therefore, it is unlikely that the increase in count rates from the
medium to the high level is caused by an localised emission region,
which is sometimes hidden from our view by a precessing accretion
disk, or by a precessing accretion disk, which sometimes blocks part
of the total X-ray radiation. \

Kuulkers et al.(1996a) proposed that the difference in overall count
rates in the Cyg-like sources is caused by anisotropic emission, the
radiation being scattered preferentially into the equatorial plane by
a puffed-up inner disk structure. In this model, with increased
scattering in the high level, the decrease in NBO amplitude follows
naturally due to light travel time effects in the scattering
process. Our observations therefore support their model.

\subsubsection{A precessing neutron star}

Another option is that the differences between the levels are due to
changes in the properties of the emission region. A precessing neutron
star could produce changes in the inner disk region. Due to the
precession of the neutron star the orientation of the magnetic field
with respect to the accretion disk changes.  According to numerical
calculations (Psaltis et al. 1995) the strength of the magnetic field
is enough for the field to have a profound effect on the spectrum of
the source and possibly also on the count rate. A change in the
effective magnetic field, caused by a different orientation of the
field, could maybe explain the different Z tracks during the different
levels. Also, the rapid X-ray variability would be affected, although
it is not clear if this could explain the differences that we
found. The behaviour of the HBO should differ between the different
intensity levels due to the different effective magnetic field, but no
difference is found. It is also not clear why the NBO
should differ between intensity levels.\

\subsection{The decrease of the HBO frequency down the NB}

The decrease of the HBO frequency from the hard vertex down the NB in
the June 1987 and October 1989 PC data allows us to apply the method
described by Wijnands et al. (1996a), in order to derive an upper
limit for the equatorial magnetic field strength. Assuming that the
radius of the neutron star is $\sim 10^6$~cm we derive an upper limit
of $\sim8.5\times 10^9$ G on the star's magnetic field at the magnetic
equator. This value is similar to the value found for GX 17+2
(Wijnands et al. 1996a). This upper limit is consistent with numerical
computations on the X-ray spectrum of Cygnus~X-2 (Psaltis et
al. 1995). As noted by Wijnands et al. (1996a), a further decrease of
the frequency down the NB will reduce this upper limit. The fact that
we see the HBO frequency decrease on the NB in Cygnus~X-2 supports the
interpretation that the QPO on the NB in GX~17+2, which also decreases
in frequency (Wijnands et al. 1996a), is the HBO.

\section{Conclusion \label{conclusion}}

We observed Cygnus~X-2 over a broad range of intensities.  The source
alternated between the known medium and high overall intensity levels
and also sometimes in a state in between these two levels, both in
intensity and other source characteristics. We found new correlations
between several characteristics of the source and the overall
intensity level :
\begin{itemize}
\item The velocity and acceleration along the normal branch
increase when the overall intensity increases.  During the high intensity
level Cygnus~X-2 moves faster and more irregularly up and down the
normal branch than during lower levels.
\item The width of the normal branch in the hardness-intensity
diagram increases when the overall intensity increases.
\item The very-low frequency noise near the soft vertex increases in
amplitude and in steepness when the overall intensity increases.
\item The normal branch quasi-periodic oscillation is not detectable
($\la$ 0.9 \% rms)
near the soft vertex when \cygx~ is in the high level. During the
other levels a NBO can be easily detected at a level of 1--2.5 \% (rms).
\end{itemize}

\noindent
The following correlations were already known and are confirmed by our
analysis :
\begin{itemize}
\item The overall intensity changes with a factor of $\sim$1.34 between the
medium and the high overall intensity level.
\item The shape of the Z-track in the colour-colour diagram and
hardness-intensity diagram changes when the overall intensity
changes. In the colour-colour diagram the horizontal and the flaring
branch become more horizontal when the overall intensity
increases. When the source enters the flaring branch during the medium
overall intensity level the intensity first increases and later
decreases. When the source enters the flaring branch during the high
overall intensity level the intensity immediately starts to decrease.
\item The whole Z-track shifts to softer colours when the overall
intensity increases.
\end{itemize}

\noindent
We also found several other previously unreported phenomena:
\begin{itemize}
\item Clear detection of overall intensity level episodes in-between the
medium and the high overall intensity level, indicating that the
different overall intensity levels are part of a continouos range
instead of a discrete set.
\item The frequency of the horizontal branch quasi-periodic
oscillation decreases down the NB, giving a model dependent upper
limit on the magnetic field strength at the magnetic equator of $\sim
8.5 \times 10^9$ G.
\item Detection of five bursts in the Ginga data, which do not show the
characteristics of bona fide type I bursts.\\
\end{itemize}

\begin{acknowledgements}This work was supported in part by the
Netherlands Organization for Scientific Research (NWO) under grant PGS
78-277 and by the Netherlands Foundation for Research in Astronomy
(ASTRON) under grant PGS 781-76-017. EK acknowledge receipts of an ESA
fellowship.\end{acknowledgements}

\appendix

\section{Calculating the standard deviation of sample standard
deviations}

\noindent
For the sample $x_1$, $x_2$, $...$, $x_n$ let

\begin{equation}
{\overline x} = {\rm mean} = \sum_{i=1}^n x_i/n
\end{equation}

\noindent
The sample standard deviation
(s) is defined by 

\begin{equation}
s = (\sum_{i=1}^n {(x_i - {\overline x})^2\over n})^{1\over2}
\end{equation}

\noindent
The standard deviation of sample standard deviations is
defined by (Burington \& May 1970) 

\begin{equation}
\sigma_s = [{n-1\over n}- [b(n)]^2]^{1\over 2}\sigma_x
\end{equation}

\noindent
with $\sigma_x^2$ the variance of the population given by the
unbiased estimate $s^2 n/(n-1)$, $b(n) = ({2\over
n})^{1\over2}\Gamma({n\over2}) / \Gamma({n-1\over 2})$ and n the
total number of points.\\

\end{document}